\documentclass[letterpaper,12pt]{article}
\pdfoutput=1

\usepackage{jheppub}
\usepackage{amsmath}
\usepackage{graphicx}
\usepackage{subfig}
\usepackage{cancel}
\usepackage[dvipsnames]{xcolor}
\usepackage{color}
\usepackage{mathrsfs}

\usepackage{amssymb}
\usepackage{epstopdf}
\def\({\left(}
\def\){\right)}
\def\[{\left[}
\def\]{\right]}

\title{Time-independent wormholes}

\author{Zicao Fu, Donald Marolf, and Eric Mefford}

\affiliation{Department of Physics, University of California, Santa Barbara, CA 93106, USA}

\emailAdd{zicaofu@physics.ucsb.edu}
\emailAdd{marolf@physics.ucsb.edu}
\emailAdd{mefford@physics.ucsb.edu}

\abstract{We study two-sided static wormholes with an exact Killing symmetry that translates both mouths of the wormhole toward the future.   This differs from the familiar Kruskal wormhole whose time translation is future-directed only in one asymptotic region and is instead past-directed in the other.  Our spacetimes are solutions to Einstein-Hilbert gravity sourced by scalar domain walls.  Explicit examples are found in the thin wall approximation.  More generally, we show that such spacetimes can arise in the presence of scalar fields with potentials that are $C^1$ but not $C^2$ and find examples numerically.  However, solutions with an exact such Killing symmetry are forbidden when the scalar potential is smooth.  Finally, we consider the mutual information of boundary regions associated with such wormholes in AdS/CFT.  Although the interior of our solutions are unstable, we find that even mutual informations between opposite boundaries are already thermalized at any finite $t$ in the sense that they agree with the $t\rightarrow \infty$ limit of results from the familiar AdS-Kruskal solution.}

\begin{document}
\maketitle

\section{Introduction}
\label{sec:Intro}

\begin{figure}[t]
\centerline{
\includegraphics[width=0.257\textwidth]{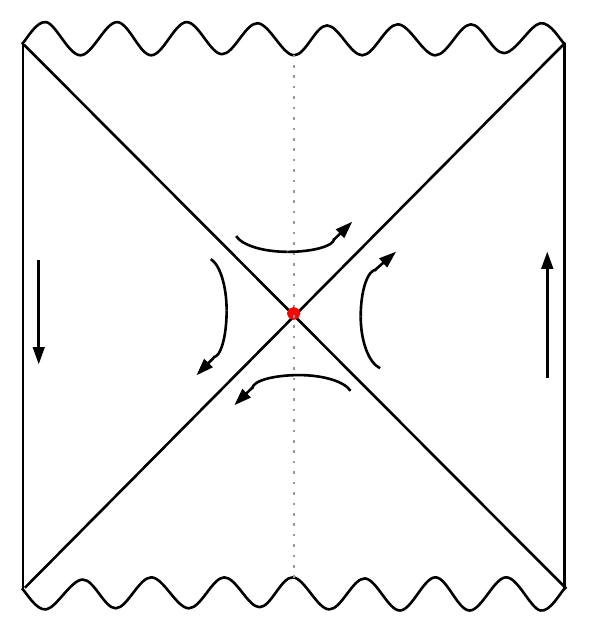}\includegraphics[width=0.5\textwidth]{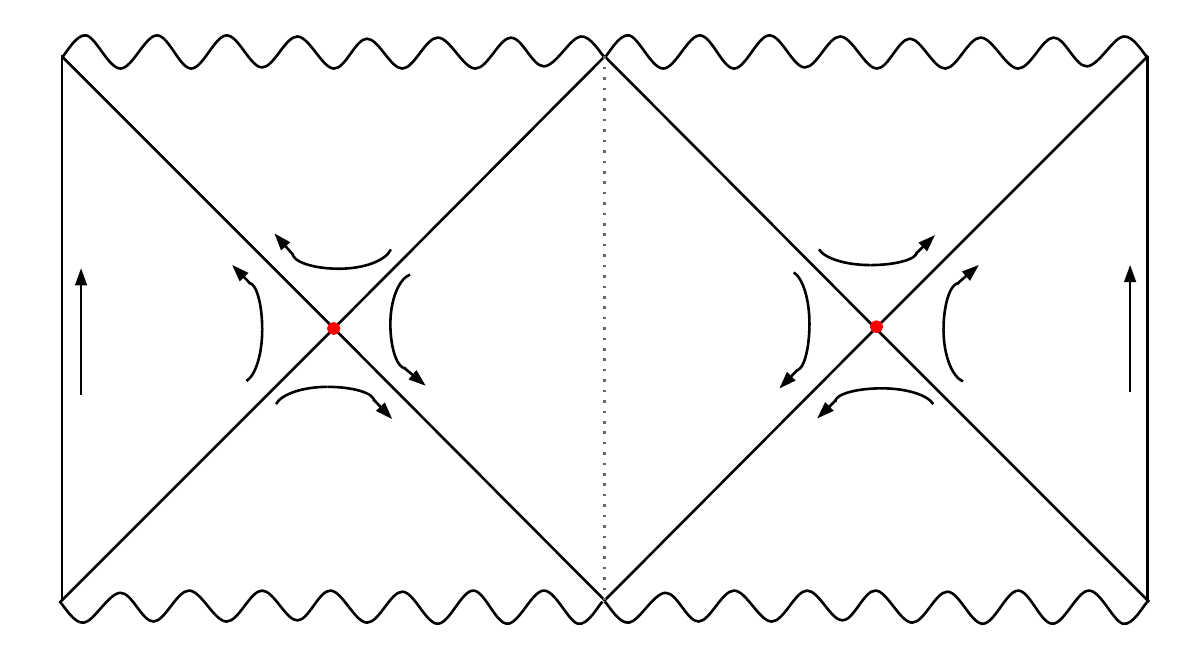}
}
\caption{Sketches of conformal diagrams for the familiar two-sided Kruskal-AdS wormhole (left) and what we call time-independent wormholes (right).  In the Kruskal case the Killing symmetry moves one boundary forward in time while shifting the other backward.  But on the right the Killing symmetry acts as a future-directed time-translation on both boundaries.  On the left, the Killing horizon has only a single bifurcation surface, while the Killing horizon of the right figure has two (red dots). Both spacetimes have $Z_2$ reflection symmetries about the dotted vertical lines. On the left, this reflection changes the sign of the time translation Killing field, while it leaves the Killing field invariant on the right.}
\label{fig:Conformal}
\end{figure}

The familiar Kruskal wormhole has an exact Killing symmetry often called a time-translation. But as illustrated in figure \ref{fig:Conformal} (left) for the asymptotically AdS case, this symmetry displaces one asymptotic region forward in time while shifting the other asymptotic region toward the past.   As a result, non-local quantities that compare the two boundaries do in fact change under the asymptotic symmetry that shifts both boundaries toward the future.  Such quantities are commonly studied in AdS/CFT and include both boundary-to-boundary two-point functions and mutual informations between the two boundaries.   The resulting time-evolutions were described in  e.g. \cite{Fidkowski:2003nf} and \cite{Hartman:2013qma}.

Below, we explore whether Einstein-Hilbert gravity coupled to familiar matter sources might allow wormholes with a Killing symmetry that translates {\it both} ends in the same direction.  Since topological censorship \cite{Friedman:1993ty,Galloway:1999bp} requires wormholes to have horizons, and since the Killing symmetry must resemble a flat-space boost transformation near the horizon bifurcation surface, such spacetimes should have conformal diagrams resembling figure \ref{fig:Conformal} (right), or more generally should have Killing horizons with an even number of bifurcation surfaces in the $t=0$ hypersurface.

For simplicity, we study wormholes with spherical symmetry.  Birkhoff's theorem then forbids vacuum solutions of this form in Einstein-Hilbert gravity.  Physically, the issue is that the interior of the wormhole tends to collapse, destroying the presumed-static region shown in the middle of the wormhole at right in figure \ref{fig:Conformal}.  We solve this problem by coupling gravity to a scalar field.  The repulsive gravity generated by either positive-tension scalar domain walls or positive scalar potentials (which effectively act as local positive cosmological constants) allow the desired static region to exist.

Section \ref{sec:Thin} constructs and studies asymptotically-AdS such solutions in the thin wall approximation. The resulting spacetimes are similar in many ways to the single-asymptotic region black holes with de Sitter interiors found in \cite{Fidkowski:2003nf}.   Interestingly, the holographic mutual information between the two boundaries always vanishes when considering regions smaller than half of either boundary.

We then consider spacetimes sourced by smooth scalar fields in section \ref{sec:Thick}.  We show that time-independent wormhole solutions exist when the scalar potential $V(\phi)$ is chosen to behave like $\phi^2 (\log \phi)^3$ near a local minimum;  i.e., while the solutions are smooth, the scalar potentials are only $C^1$ as functions of $\phi$.  Examples are constructed numerically.   Again, the holographic mutual information between the two boundaries always vanishes when considering regions smaller than half of either boundary. That singular potentials are required is shown in section \ref{sec:NoGo}; scalar fields with smooth potentials cannot support our time-independent wormholes.  We close with some final discussion in section \ref{sec:Disc}.  In particular, we comment on the status of such solutions with respect to gauge/gravity duality and also with respect to recent discussions of the possible role of complexity in gauge/gravity duality \cite{Stanford:2014jda,Roberts:2014isa,Brown:2015bva,Brown:2015lvg}.

\section{Thin Wall Solutions}
\label{sec:Thin}
We begin in section \ref{sec:cut} by constructing thin wall versions of the time-independent wormholes shown at right in figure \ref{fig:Conformal}.  We then briefly analyze the holographic mutual information defined by these wormholes in section \ref{sec:MIthin} and note that in a certain sense they are already thermalized at any finite $t$.

\subsection{A cut and paste construction}
\label{sec:cut}

It is straightforward to assemble the desired time-independent wormholes by cutting two copies of Kruskal-AdS (fig. \ref{fig:Conformal} left) along a timelike surface defined by orbit of the symmetry group (a constant $r$ surface) and then sewing the two larger pieces together other along a thin positive-tension domain wall. This domain wall then becomes the dotted line in right diagram in figure \ref{fig:Conformal} and is left invariant under the reflection symmetry.

To proceed, recall the $D$ dimensional AdS-Schwarzschild metric
\begin{equation}
ds^2=-\left(1-\frac{\omega _DM}{r^{D-3}}+\frac{r^2}{\ell ^2}\right)dt^2+\frac{1}{1-\frac{\omega _DM}{r^{D-3}}+\frac{r^2}{\ell ^2}}dr^2+r^2d\Omega ^2,
\end{equation}
where $\omega _D=\frac{16\pi G_D}{\left(D-2\right)S_{D-2}}$ and $S_{D-2}=\frac{2\pi ^\frac{D-1}{2}}{\Gamma \left(\frac{D-1}{2}\right)}$.  A timelike constant $r$ surface has unit normal $n^a=\sqrt{1-\frac{\omega _DM}{r^{D-3}}+\frac{r^2}{\ell ^2}}\left(\frac{\partial }{\partial r}\right)^a$. Its extrinsic curvature $K_{ab}=\frac{1}{2}\pounds _nh_{ab}$ is thus
\begin{equation}
\label{eq:ExtrCurv}
K_{ab} dx^a dx^b=\frac{1}{2}\sqrt{1-\frac{\omega _DM}{r^{D-3}}+\frac{r^2}{\ell ^2}}\left[-\left((D-3)\frac{\omega _DM}{r^{D-2}}+\frac{2r}{\ell ^2}\right)dt^2 +2rd\Omega^2\right].
\end{equation}

We wish to consider relativistic domain walls with surface stress tensor $\hat T_{ab}=-\sigma h_{ab}$ in terms of the (constant) tension $\sigma$ and the induced metric $h_{ab}$.  Here we use the conventions of \cite{Wald:1984rg} in which $h_{ab}$ is a degenerate tensor in the full spacetime such that $h^a_{~b}$ is the projector onto the vector space tangent to the wall.  The full stress-energy tensor $T_{ab}$ is proportional to $\hat T_{ab}$, but contains an extra delta-function localizing the stress-energy on the wall.  Given the $Z_2 $ symmetry of figure \ref{fig:Conformal} (right), the Israel junction conditions (see e.g. \cite{Misner:1974qy}) require $\hat T_{ab} \propto K_{ab}$, and thus $\frac{g_{tt}}{g_{\Omega \Omega }}=\frac{K_{tt}}{K_{\Omega \Omega }}$. This relation is satisfied if and only if
\begin{equation}
\label{eq:sew}
r_\text{wall}^{D-3}=\frac{D-1}{2}\omega _DM.
\end{equation}
The junction condition then gives $K_{ab} = \frac{4\pi G_D\sigma }{D-2}h_{ab}$ so that
\begin{equation}
\sigma = \frac{D-2}{4\pi G_Dr_\text{wall}}\sqrt{1 - \frac{\omega_D M}{r_\text{wall}^{D-3}} + \frac{r_\text{wall}^2}{\ell ^2}}
\end{equation}
is positive as desired.

This completes our construction of thin-wall solutions corresponding to figure \ref{fig:Conformal} (right).  However, we note in passing that a similar analysis indicates that our solutions are unstable.  This is to be expected as the interior of our wormhole remains static only due to a delicate balance between the gravitational attraction of the black hole and the gravitational repulsion of the domain wall.  Indeed, maintaining the $Z_2$ reflection symmetry and spherical symmetry but allowing the wall to move with time on a surface $r = R(T)$, the Israel junction conditions imply an equation of motion

\begin{equation}
\label{eq:tIJC}
2\sqrt{f(R)+\dot R^2}=\frac{8\pi G_D\sigma }{D-2}R,
\end{equation}
for $f(r)=1-\frac{\omega _DM}{r^{D-3}}+\frac{r^2}{\ell ^2}$ and $\dot R$ the derivative of $R$ with respect to proper time along the shell.  Here the first-order nature of the equation is a consequence of restricting to solutions with $Z_2$ symmetry. Squaring \eqref{eq:tIJC} and linearizing it around the static solution \eqref{eq:sew}, we obtain

\begin{equation}
\left(\frac{d}{d\tau }\delta R\right)^2=\left(D-3\right)4^\frac{1}{D-3}\left(\left(D-1\right)\omega _DM\right)^\frac{-2}{D-3}\delta R^2+O(\delta R^3),
\end{equation}
so the static solution is unstable on the timescale

\begin{equation}
\tau =\sqrt{\frac{1}{D-3}4^\frac{-1}{D-3}\left(\left(D-1\right)\omega _DM\right)^{\frac{2}{D-3}}}.
\end{equation}

\subsection{Mutual Information and Thermalization}
\label{sec:MIthin}

As noted in the introduction, physical quantities defined by the geometry of our wormhole must be independent of time.  This includes the (leading order) holographic mutual information defined by the Ryu-Takayanagi (RT) \cite{Ryu:2006bv,Ryu:2006ef} or the covariant Hubeny-Rangamani-Takayanagi (HRT) \cite{Hubeny:2007xt} prescriptions.  While -- as will be discussed in section \ref{sec:Disc} -- the derivations of \cite{Lewkowycz:2013nqa} and \cite{Dong:2016hjy} need not apply to our spacetime, it is nevertheless of interest to investigate what these prescriptions would predict.  In particular, we will see that -- despite the instability noted above -- in a sense these mutual informations (and indeed the entropies of all boundary regions) appear to already be thermalized at any finite $t$.

We note that such leading-order holographic mutual informations are of more interest in our context than our boundary-to-boundary correlators, as the latter depend on the choice of quantum state for light bulk fields as well as on the classical background geometry.  Since we have not constructed our spacetimes as stationary points of a path integral, there is no preferred choice for this quantum state.  And due to the large causal shadow between the two event horizons of our time-independent wormholes, we are free to choose the light bulk fields in the left asymptotic region to be completely uncorrelated with those in the right asymptotic region so that all connected correlators vanish when evaluated with one argument on the right boundary and another on the left.

Because the spacetime is not globally static, the RT prescription does not strictly apply.  Nevertheless, in a spacetime with time-reversal symmetry, the maximin construction of \cite{Wall:2012uf} guarantees the HRT surface to be the minimal surface within the $t=0$ (i.e., within the hypersurface invariant under $t \rightarrow -t$) as one would expect from the RT prescription\footnote{\label{VH}Since this surface is minimal on the $t=0$ slice, its area can be no larger than that of the maximin surface. But the time-reversal symmetry means that this minimal surface is also an
extremal surface in the full spacetime. It can therefore have area no smaller than the maximin surface,
as the latter agrees with the area of the smallest extremal surface. We thank Veronika Hubeny for pointing this out to us.}.

We wish to study surfaces anchored both to a region $A_R$ of the right boundary and also to a corresponding region $A_L$ of the left boundary, such that $A_R, A_L$ are interchanged by the $Z_2$ symmetry of reflection across the wall. In order to compute the entropy $S_{A_LA_R}$ of $A_L \cup A_R$, we must correctly identify the minimal surface.  We first consider the case where $A_R$ and $A_L$ are each precisely half of the $t=0$ sphere at the AdS boundary (note that our solutions correspond to `global' Schwarzschild-AdS). Referring to $A_L, A_R$ as the `northern' hemispheres (whose boundaries are thus the equator of the sphere), it is then clear that the smallest connected surface anchored to both $A_L$ and $A_R$ is the surface defined by taking the equator of the sphere at each $r$.  As shown below, it suffices for our purposes to compute the area of the portion of this surface inside our wormhole.  Noting that the radius $r_{0}$ of the event horizon is defined by

\begin{equation}
1-\frac{\omega _DM}{r_{0}^{D-3}}+\frac{r_{0}^2}{\ell ^2} =0,
\end{equation}
and introducing $\tilde r = r/\ell$ and $\tilde r_{0} = r_{0}/\ell$,
this area satisfies

\begin{equation}
\label{eq:extrathin}
\frac{A_\text{connected, \ inside}}{A_\text{EH}}=\frac{2}{\sqrt{\pi }}\frac{\Gamma \left(\frac{D-1}{2}\right)}{\Gamma \left(\frac{D-2}{2}\right)}\int_{1}^{\left[\frac{D-1}{2}\left(1+\tilde r_{0}^2\right)\right]^{\frac{1}{D-3}}}\frac{\hat{r}^{D-3}}{\sqrt{1-\frac{1+\tilde r_{0}^2}{\hat{r}^{D-3}}+\left(\tilde r_{0} \hat{r}\right)^2}}d\hat{r},
\end{equation}
where we have normalized the quantity by dividing by the area of either event horizon.  The area of the full minimal connected surface is then $A_\text{connected} = A_\text{connected, \ inside} + A_\text{connected, \ Kruskal}$ where $A_\text{connected, \ Kruskal}$ is the area of the minimal connected surface in the AdS-Kruskal geometry of figure \ref{fig:Conformal} (left).

\begin{figure}[t]
\centering
\includegraphics[width=0.9\textwidth]{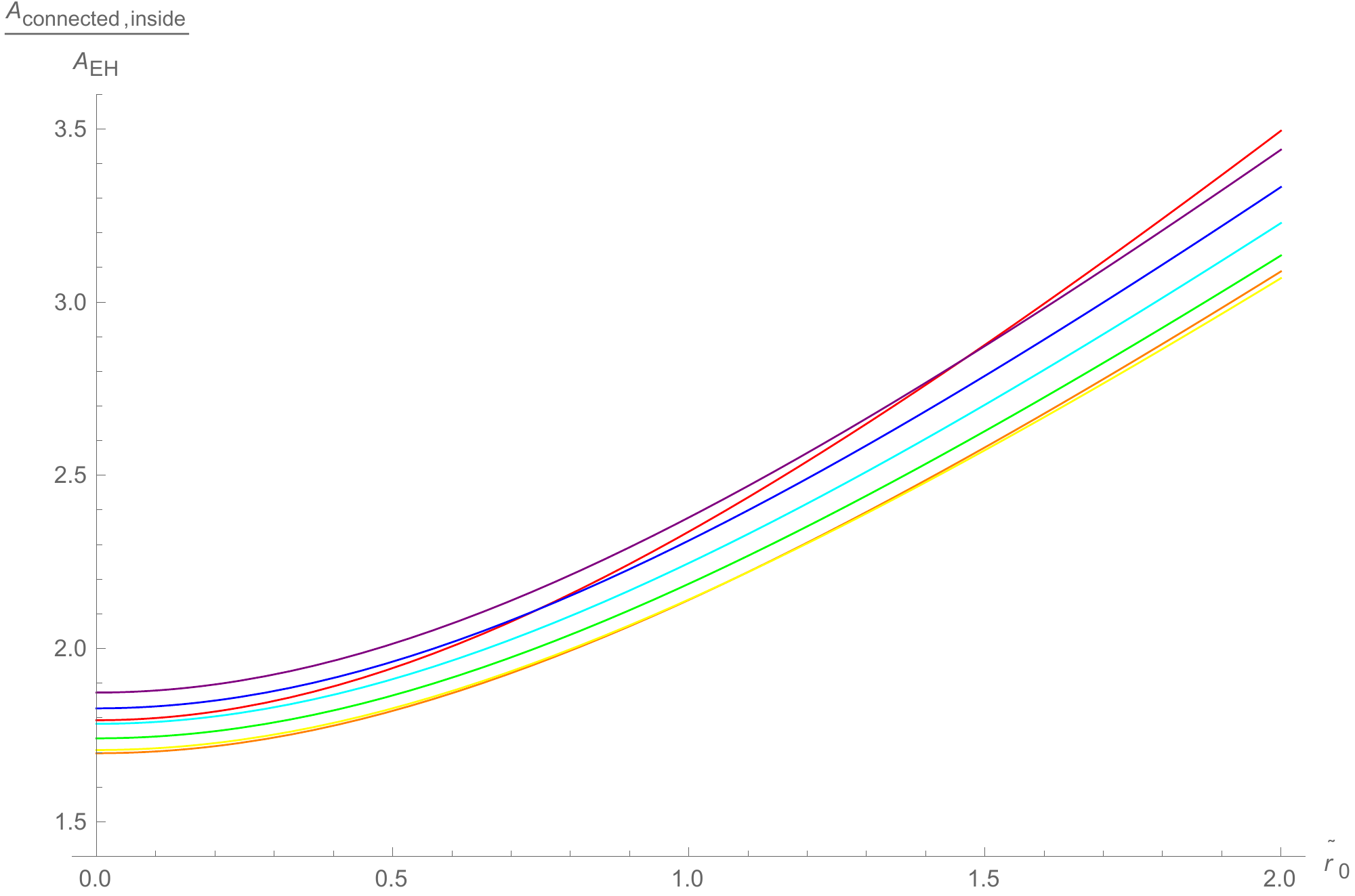}
\caption{The functions \eqref{eq:extrathin} for $D=4$ (red), $D=5$ (orange), $D=6$ (yellow), $D=7$ (green), $D=8$ (cyan), $D=9$ (blue), and $D=10$ (purple).}
\label{fig: MutualInfo}
\end{figure}

For general $D$ the integral \eqref{eq:extrathin} can be performed numerically.  But for $D=5$ it can be performed exactly to obtain
\begin{equation}
\label{eq:extrathin5d}
\frac{A_{\text{connected, inside, }D=5}}{A_\text{EH}}=\frac{2{{\tilde r}_{0}}\sqrt{1+{\tilde r}_{0}^{2}}\left( 1+2{\tilde r}_{0}^{2} \right)-\ln \left( 1+2{{\tilde r}_{0}}\left( {{\tilde r}_{0}}+\sqrt{1+{\tilde r}_{0}^{2}} \right) \right)}{\pi {\tilde r}_{0}^{3}}.
\end{equation}
As shown in figure \ref{fig: MutualInfo}, \eqref{eq:extrathin} and \eqref{eq:extrathin5d} are increasing functions of $\tilde r_0$, which are larger than $1.6$ for all $\tilde r_0$ (at least for $4 \le D \le 10$). In particular, there is $\frac{A_\text{connected, inside}}{A_\text{EH}}>1$.

\begin{figure}[t]
\centering
\includegraphics[scale=.4]{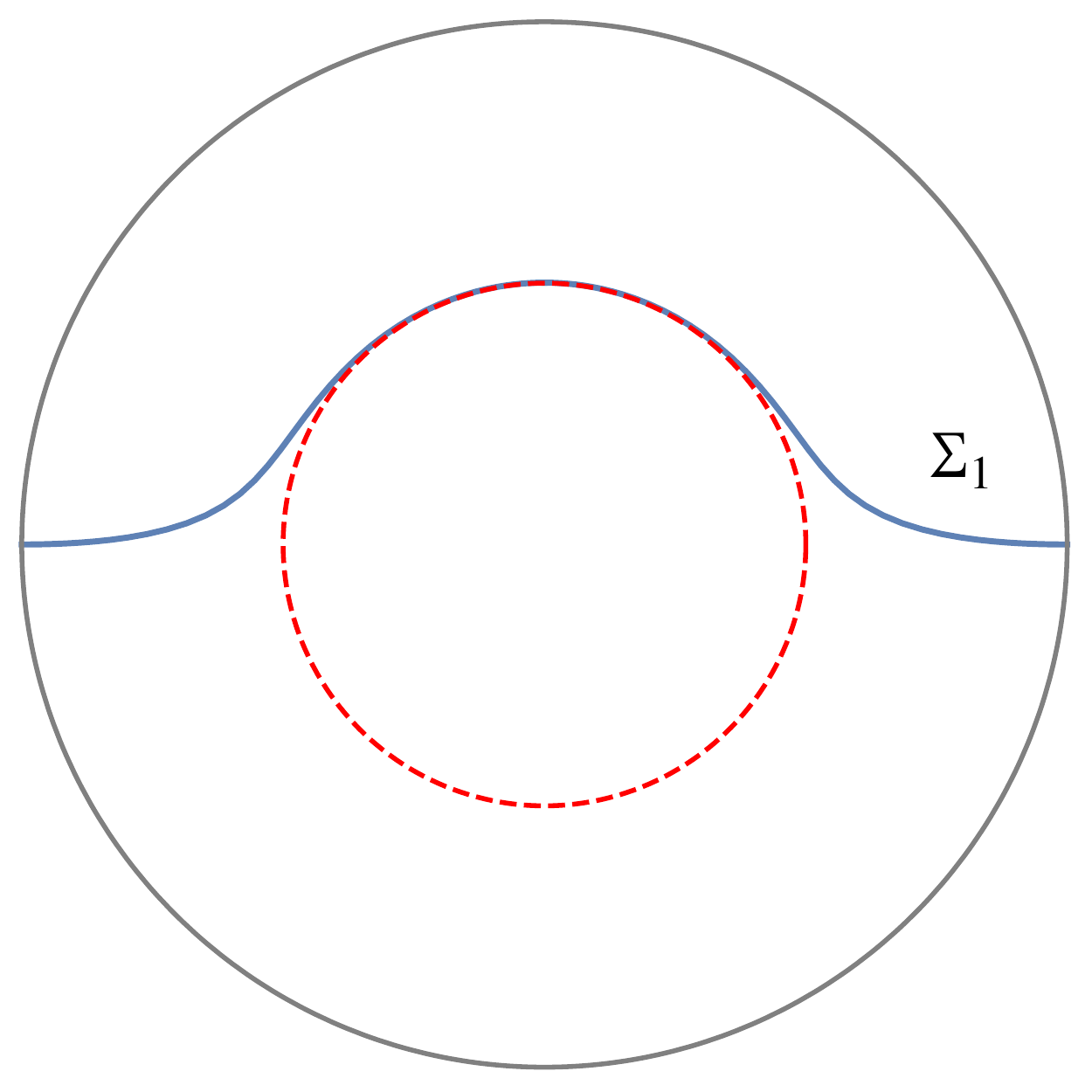}
\includegraphics[scale=.4]{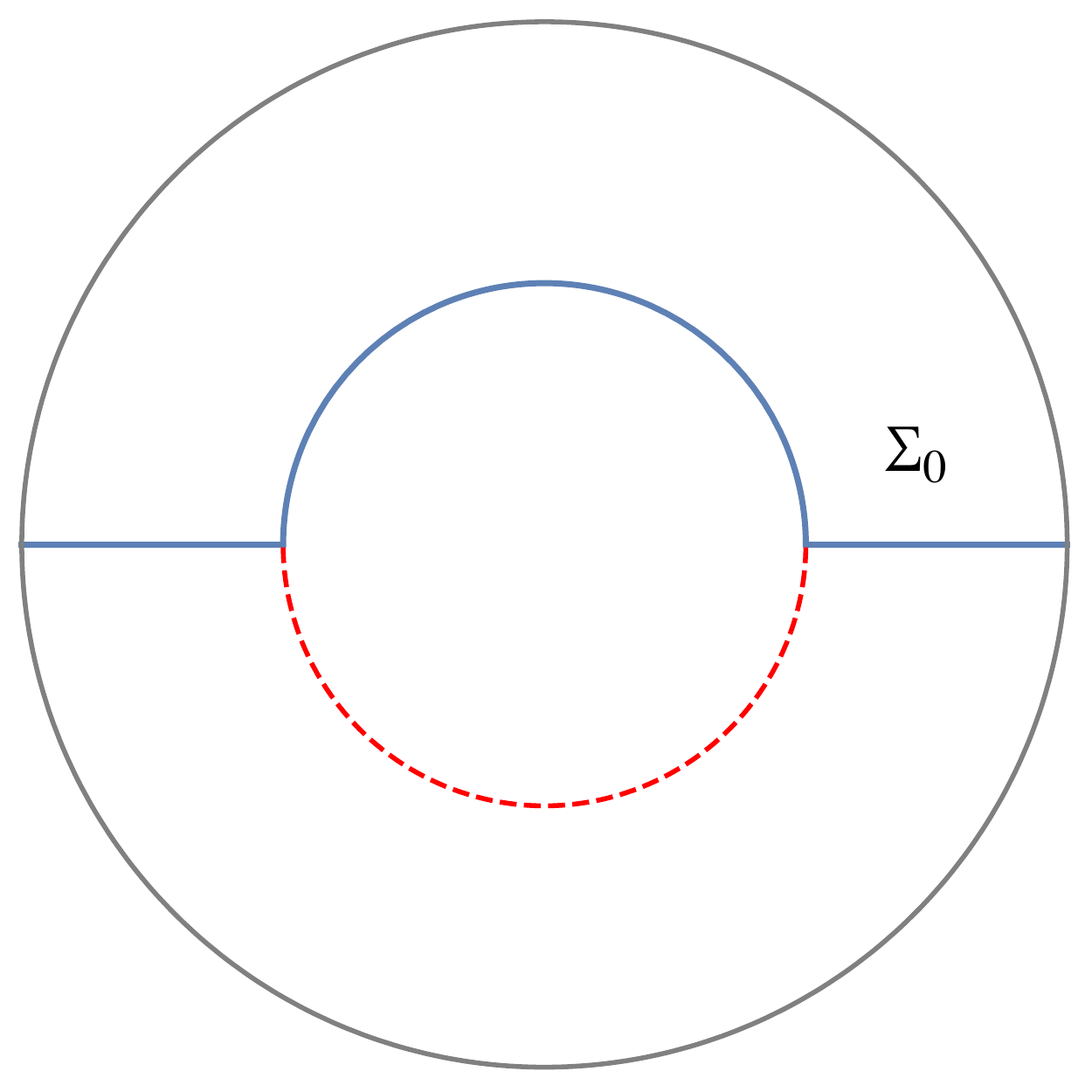}
\caption{On the left, $\Sigma_1$ is minimal surface for a hemisphere of the boundary with a black hole (red, dotted) in the bulk. The surface $\Sigma_0$ on the right necessarily has larger area than $\Sigma_1$. This surface contains a piece (straight segments along the equator) that are part of the connected surface passing through the wormhole; the other piece lies on the black hole horizon.}
\label{fig:figure1}
\end{figure}

However, as usual we must also consider the smallest disconnected surface anchored on $A_L, A_R$ and compare its area to that of the connected surface. Let us first study a single connected component, say the one anchored to $A_L$.  One example of a surface satisfying these boundary conditions is the surface $\Sigma_0$ shown in figure \ref{fig:figure1} which consists of the northern hemisphere of the bifurcation surface for the left event horizon together with the equators of all $t=0$ spheres in the left asymptotic region.   In other words, outside the horizon it coincides with the connected surface studied above anchored to both $A_L$ and $A_R$.  So the area of the left component of the actual minimal surface must be less than that of $\Sigma_0$.

Adding together the two components, the area of the minimal disconnected surface must satisfy
\begin{equation} A_\text{disconnected} \le   A_\text{connected, \ Kruskal} +  A_\text{EH}.
\end{equation}
The observation that
\eqref{eq:extrathin} and \eqref{eq:extrathin5d} are larger than $1$ then implies
$A_\text{disconnected} < A_\text{connected}$.  The HRT surface is thus disconnected and, due to e.g. the barrier theorems of \cite{Engelhardt:2013tra}, lies entirely outside the horizons.  The mutual information $I(A_L:A_R)$ is then just what would be obtained from surfaces outside the horizon of AdS-Kruskal (fig. \ref{fig:Conformal} left) and $I(A_L:A_R)$ vanishes.  Furthermore, the positivity and monotonicity of HRT mutual information derived in \cite{Wall:2012uf} then imply vanishing mutual information
$I(A_L:A_R)$ for any subsets $A_L, A_R$ of the northern hemisphere, whether or not such $A_L,A_R$ are related by the $Z_2$ symmetry.

In fact, since $A_L \cup A_R$ is homologous to its complement, the same argument shows that the HRT surface for $S_{A_LA_R}$ is again disconnected (and lies entirely outside the horizon) whenever $A_L, A_R$ both {\it contain} the entire southern hemisphere.  So here too $I(A_L:A_R)$ is what would be obtained from surfaces outside the horizon of AdS-Kruskal (fig. \ref{fig:Conformal} left), though due to the homology constraint $I(A_L:A_R)$ no longer vanishes.

Equivalently \cite{Hartman:2013qma}, we may say in both cases that $I(A_L:A_R)$ for the time-dependent wormhole agrees with that for the $t \rightarrow +\infty$ limit of AdS-Kruskal.  Though there remain certain cases that we have not checked, it is thus natural to conjecture the same to be true of arbitrary $A_L, A_R$, and thus for the entropies of arbitrary boundary regions.  But the $t\rightarrow +\infty$ limit of AdS-Kruskal is naturally interpreted as a thermalized state.  So if our conjecture is true, then despite the instability found in section \ref{sec:cut}, as measured by such entropies we find that our time-independent wormhole is already thermalized at any finite time $t$.

\section{Smooth Solutions}
\label{sec:Thick}

Having constructed time-independent wormholes using thin shells, it is natural to ask if similar solutions can be sourced by smooth scalar fields $\phi$.  We shall now show that they can, but with an interesting twist.  While the solutions are completely smooth, the scalar potential $V(\phi)$ is not.  Indeed, near the AdS minimum $\phi_0$, our $V(\phi)$ will behave like $(\phi-\phi_0)^2 [\ln(\phi-\phi_0)]^3.$  We demonstrate the existence of such solutions analytically and construct a particular example numerically.  Appendix \ref{sec:NoGo} then gives a general argument that spherically symmetric time-independent wormholes cannot be sourced by scalar fields with smooth potentials.

Our smooth solution will bear a strong similarity to our domain wall solution, in that it will be precisely $D$-dimensional AdS-Schwarzschild outside the horizon and also in the region where the time-translation Killing field is spacelike.  In those regions our scalar field will be constant and will sit at a minimum of its potential.  The scalar will deviate from this minimum only in the central diamond of figure \ref{fig:Conformal} (right) which in section \ref{sec:Thin} contained the domain wall; we refer to this diamond as the wormhole below.  Smoothness then requires that all derivatives of $\phi$ vanish at boundaries of the wormhole.

The wormhole should enjoy both spherical and time-translation symmetry.  As a result, any smooth metric in this region may be written

\begin{equation}
\label{eq:wmet}
ds^2 = -f(r)dt^2 + \frac{dr^2}{f(r)} + S(r)^2d\Omega_{D-2}^2,
\end{equation}
where we require $f$ to vanish linearly at the wormhole boundaries to give a smooth bifurcate horizon.
Imposing the $Z_2$ reflection symmetry of figure \ref{fig:Conformal} (right), we may set $r=0$ at the fixed points of this reflection.  It then suffices to study the metric only on the right half of the spacetime. We take this to be $r > 0$, with the wormhole boundary at $r=r_h$. Note that \eqref{eq:wmet} and these choices still allow the freedom to perform a constant rescaling $(t,r,r_h,f) \rightarrow (\alpha t, r/\alpha , r_h/\alpha, f/\alpha^{2})$ without changing the geometry. For later reference, we note that
AdS-Schwarzschild in these coordinates has
\begin{equation}
S_{AS}(r) = r,\quad f_{AS}(r) = \frac{r^2}{\ell^2} + 1 - (r_h^2+\ell^2)\left(\frac{r_h}{r}\right)^{D-3},
\label{eq:AdSSchwarz}
\end{equation}
with AdS boundary at $r\to \infty$. From now on, we set $\ell=1$ so that
\begin{equation}
\label{eq:AdSmin}
V(\phi(r_h)) = \Lambda_{AdS} = -\frac{(D-1)(D-2)}{2}.
\end{equation}

In these coordinates, the equation of motion for a single minimally coupled scalar field reads
\begin{equation}
f\phi''+\left[(D-2)\frac{fS'}{S}+f'\right]\phi' = \frac{dV}{d \phi}.
\label{eq:scalar}
\end{equation}
and the nontrivial $tt$, $rr$, and sphere-sphere, components of the Einstein equation (with $8\pi G=1$) may be combined to write
\begin{equation}
\begin{split}
(D-2)S''&=-S\phi'^2, \\
\left(\frac{S'}{S}\right)f' -\frac{D-3}{S^2}\left(1-f(S')^2\right) &= \frac{2}{D-2}T_r^r = \frac{1}{D-2}(f\phi'^2-2V(\phi)),\\
f'' +(D-3)\left(\left(\frac{S'}{S}\right)f'+\frac{1}{S^2}-\left(\frac{S'^2}{S^2}\right)f\right) &= \frac{2}{D-2}T_\Theta^\Theta =-\frac{1}{D-2}\left(2V(\phi)+f\phi'(r)^2\right).
\label{eq:Einstein}
\end{split}
\end{equation}
As usual, \eqref{eq:scalar} follows from \eqref{eq:Einstein} due to the Bianchi identity, so it suffices to consider \eqref{eq:Einstein}.

Rather than choose a form for $V(\phi)$ and solve for the resulting $\phi(r)$, we find it convenient to proceed in analogy with section 4 of \cite{Herdeiro:2015waa} and to posit $\phi(r)$.  We then take the middle equation from \eqref{eq:Einstein} as the definition of $V(\phi)$.  The requirement that all derivatives of $\phi(r)$ vanish at $r_h$ motivates us to choose the form
\begin{equation}
\phi(r) = b\tanh\left(\frac{kr}{r_h^2-r^2}\right).
\label{eq:scalarprofile}
\end{equation}

This leaves us with a pair of second order ODEs (the first and last of \eqref{eq:Einstein}) to solve for $f(r), S(r)$. The $Z_2$ reflection symmetry requires the boundary conditions
\begin{equation}
S'(0) = f'(0) = 0.
\label{eq:midpoint}
\end{equation}
We also wish to impose two boundary conditions at $r_h$.  The first of these is simply $f(r_h)=0$.  Using \eqref{eq:AdSmin} and our definition of $V(\phi)$ (the middle equation in \eqref{eq:Einstein}) gives the second:
\begin{equation}
\label{eq:hBC2}
\frac{df}{dr}|_{r=r_h} = \frac{1}{S'(r_h)S(r_h)}\left((D-1)\;S(r_h)^2 + (D-3)\right).
\end{equation}

We note that \eqref{eq:hBC2} guarantees the surface gravity at the wormhole boundary to match that of AdS-Schwarzschild with horizon radius $S(r_h)$ if we rescale
$(t,r,r_h,k,f) \rightarrow (\alpha t, r/\alpha , r_h/\alpha, \alpha k, f/\alpha^{2})$ to set $S'(r_h)=1.$   With this understanding, the redshift factor $f$, the sphere size $S$, and their first derivatives with respect to $r$ are then continuous at $r=r_h$.  So long as $S(r_h) \neq 0$, the ODEs \eqref{eq:Einstein} then guarantee continuity of all derivatives and the geometry matches smoothly to AdS-Schwarzschild as desired.  However, it will be convenient for our later numerics to first choose $b,k,r_h$ arbitrarily and only later to rescale in this manner.

\begin{figure}[t]
\centering
\includegraphics[scale=.3]{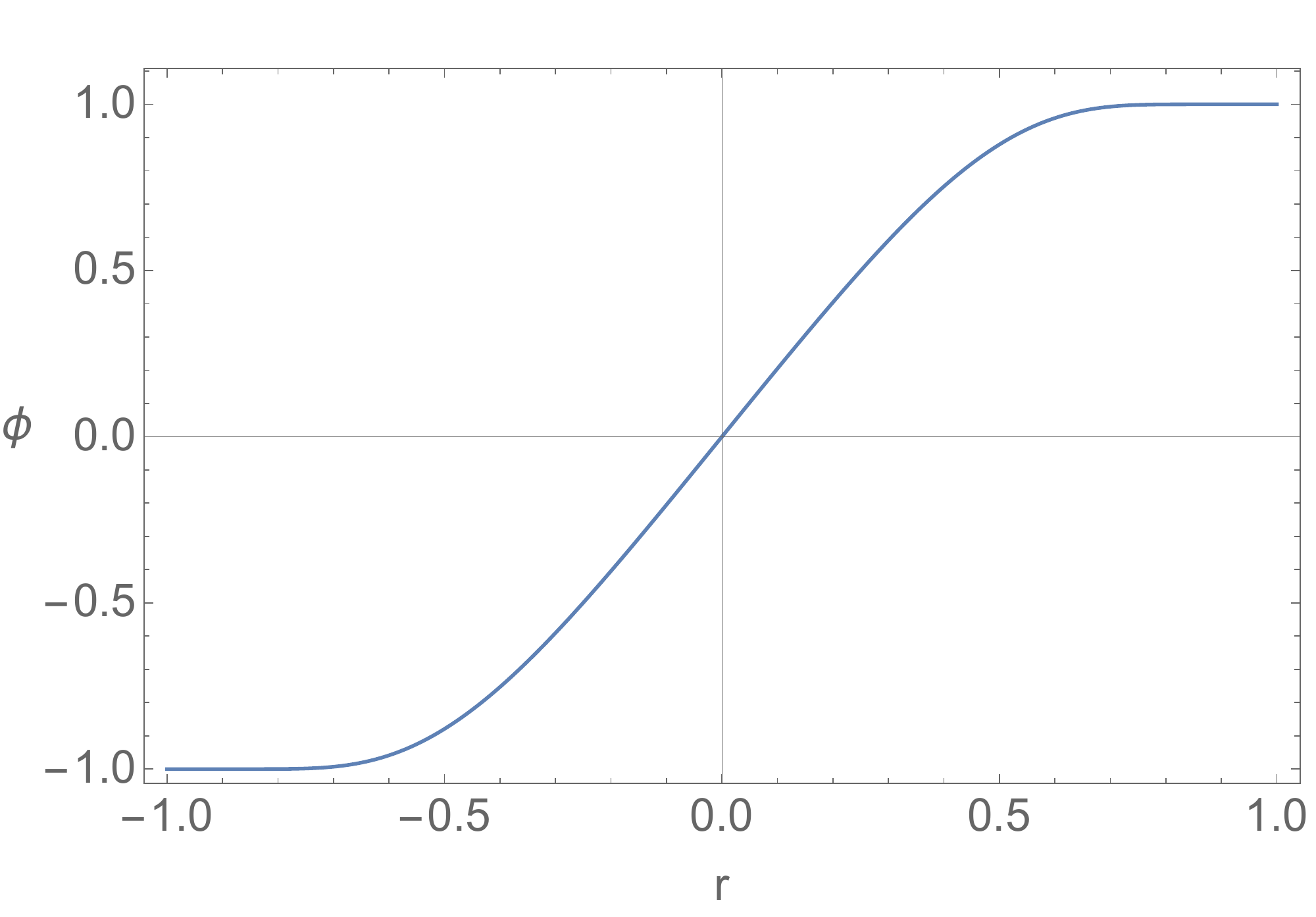}
\includegraphics[scale=.3]{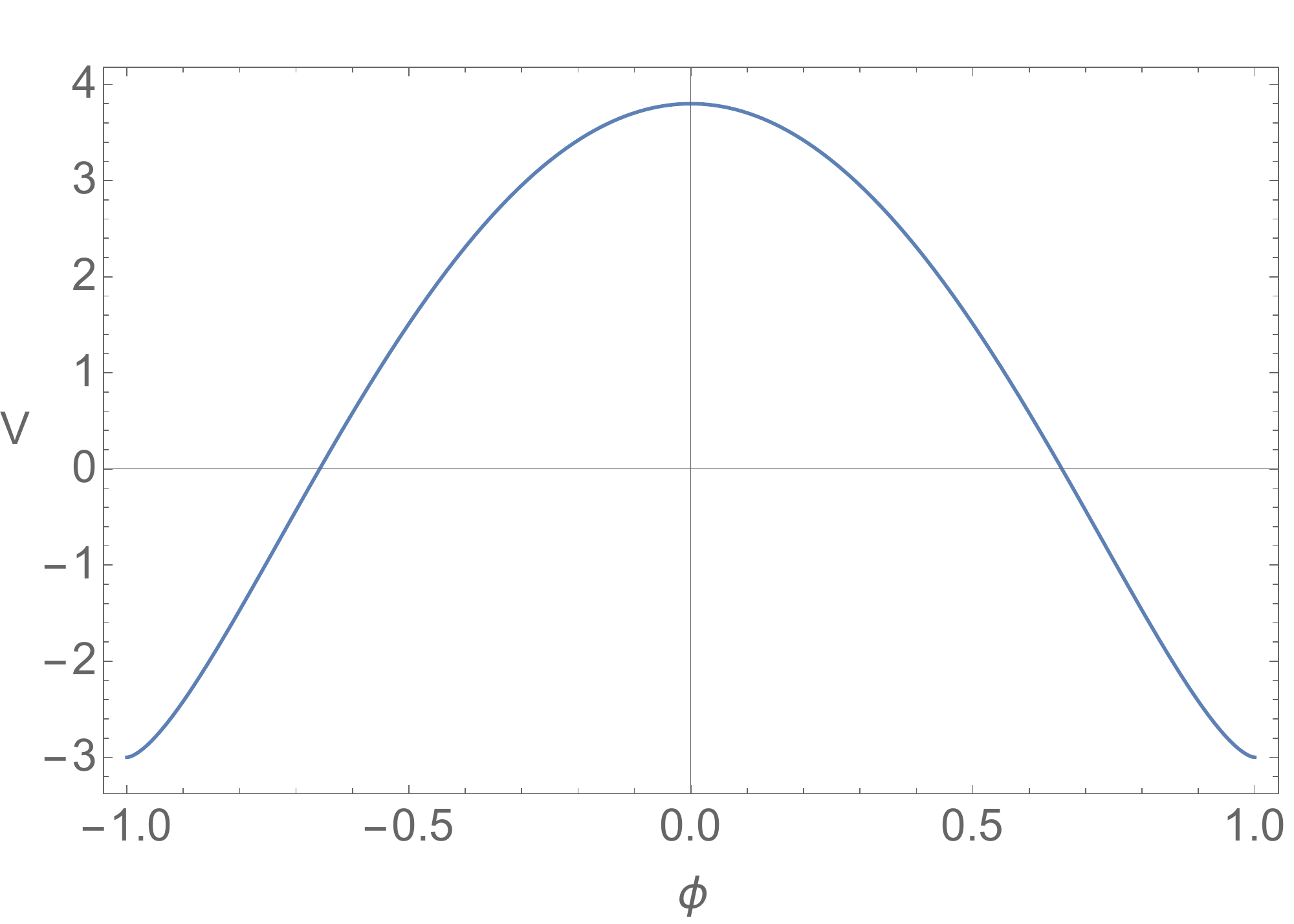}
\centering
\includegraphics[scale=.3]{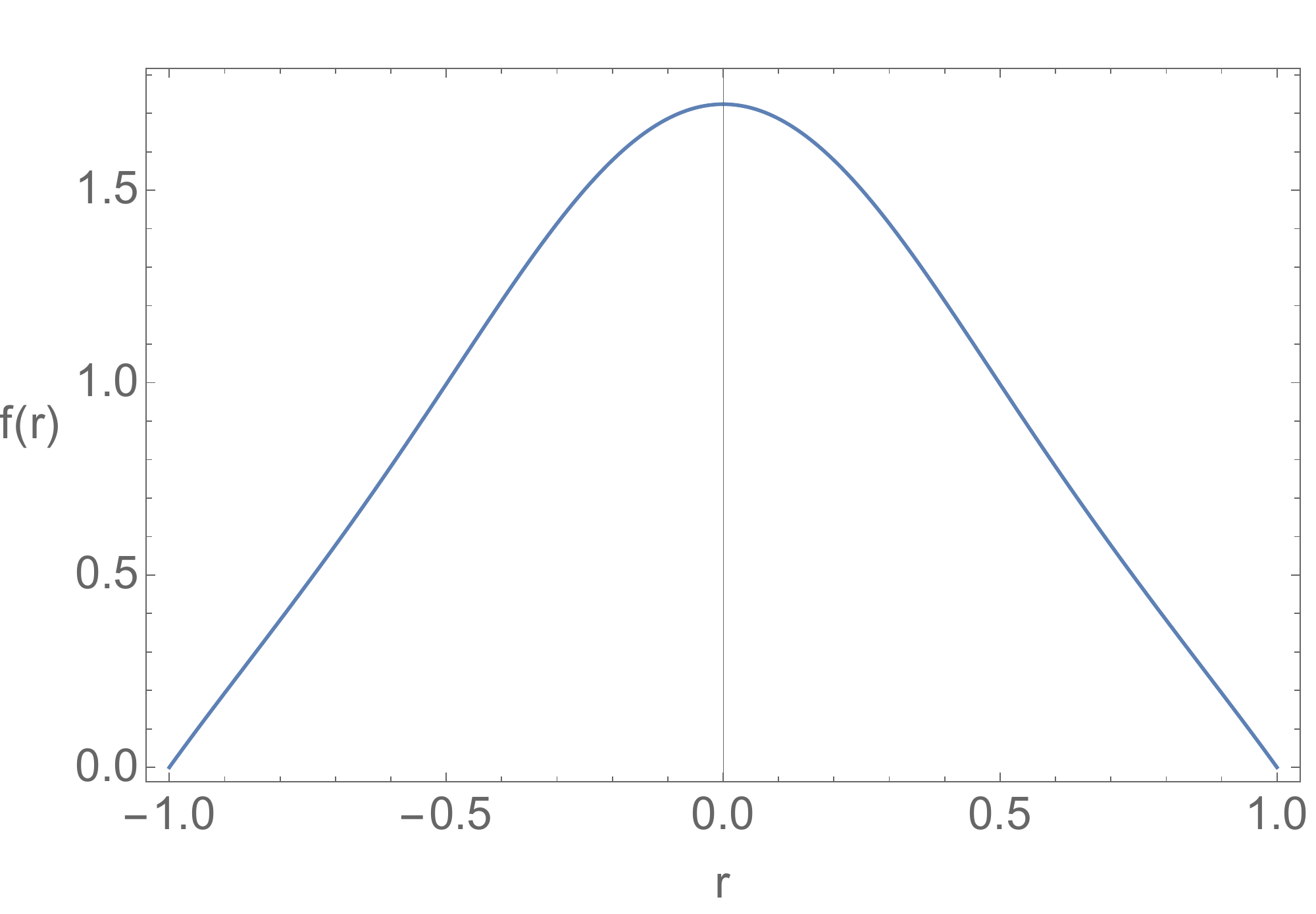}
\includegraphics[scale=.3]{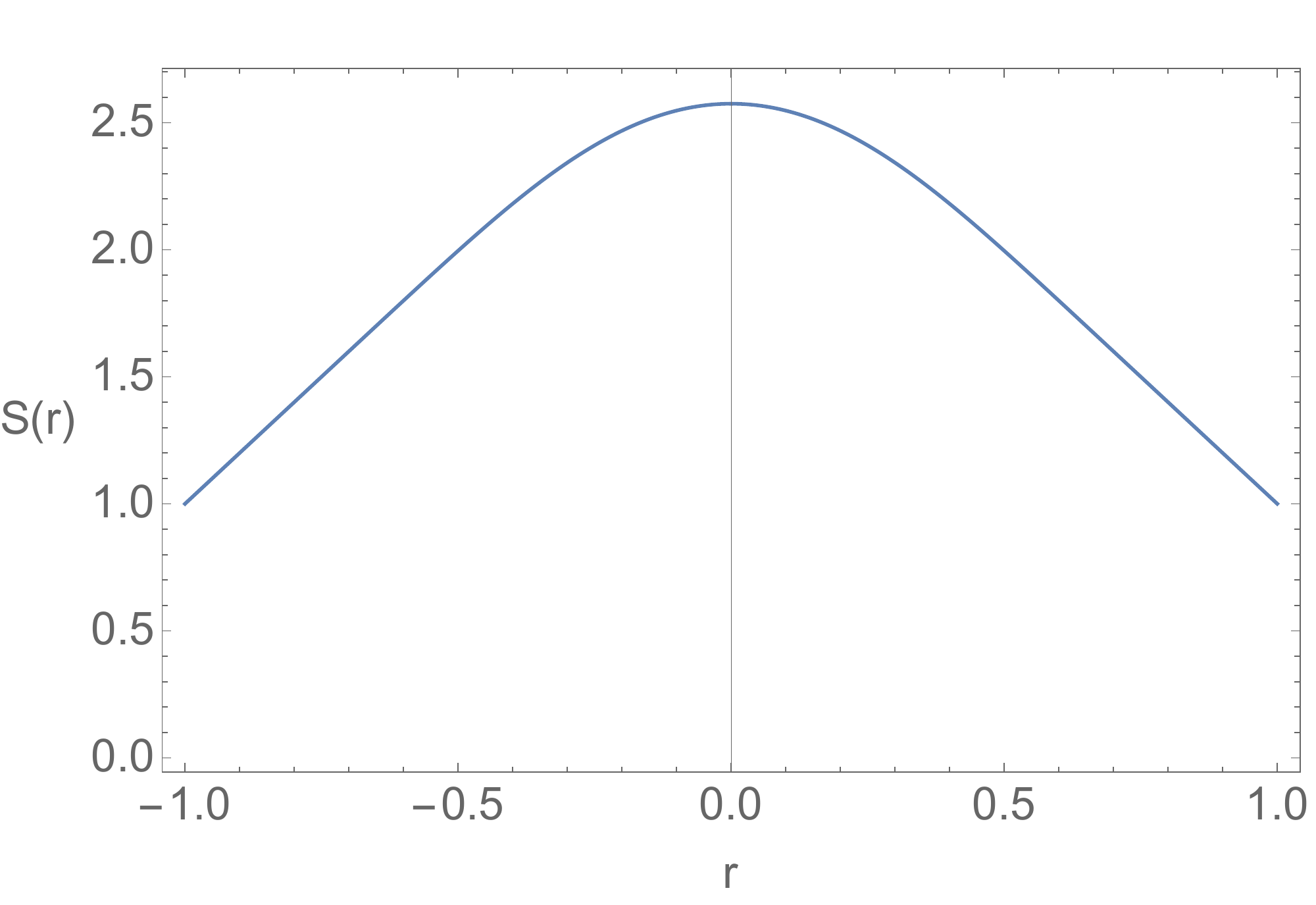}
\caption{Above we plot the numerical solutions to the Einstein-scalar system for the scalar field profile (\ref{eq:scalarprofile}) with b = 1, k=2.05768. Note that, having set $\ell=1$, the potential goes to $V=\Lambda_{AdS}$ at the horizons ($r_h=\pm 1$).}
\label{fig:solutions}
\end{figure}

This suffices to prove that the desired solutions exist. So long as $S> 0$, it is clear that our ODEs have no singular points.  Furthermore, since we take $\phi(r)$ as given, the first ODE is a homogeneous equation for $S(r)$ alone.  Using only $S'(0)=0$, it is then clear that the resulting one-parameter family of solutions for $S(r)$ will have $S > 0$ on $[0,r_h]$ so long as we choose $b$ sufficiently small for given $k, r_h$.  For each such $S(r)$, the second ODE defines a regular linear 2nd order ODE for $f(r)$, so there a unique solution $f(r)$ satisfying $f'(0)=0$ and $f(r_h)=0$.  Taking the remaining free parameter to be $S(0)$, and noting that scaling $S(0) \rightarrow \beta S(0)$ induces the scalings $S(r), f(r), V(r) \rightarrow \beta S(r), \beta^{-2} f(r), \beta^{-2} V(r)$ we may then choose $\beta$ so as to both satisfy \eqref{eq:hBC2} and make $S(0)$ positive.  Thus smooth time-independent wormholes of this form exist so long as $b$ is sufficiently small.  Figure \ref{fig:solutions} displays numerical solutions for $f(r)$, $S(r)$ and $V(\phi)$ in $D=4$ with
\begin{equation}
b = 1, \quad k = 2.05768, \quad S(r_h)=1,
\end{equation}
where for numerical convenience we have chosen $r_h=1$.

It now remains to discuss $V(\phi)$.  Since $f, S$ are smooth, our definition of $V$ via \eqref{eq:Einstein} guarantees that $V$ is a smooth function of $r$.  The ansatz \eqref{eq:scalarprofile} then implies that $V(\phi)$ is smooth for $\phi \in (-b,b)$.  But the behavior at the minimum $b$ must be determined by expanding $S,f,\phi$ near $r=r_h$.   To simplify this calculation we now set $r_h=1$ to find
\begin{equation}
\begin{split}
\phi &\approx b(1-2e^{-k/(1-r)}),\\
\phi' &= b\left(\frac{kr}{1-r^2}\right)'\text{sech}^2\left(\frac{kr}{1-r^2}\right) \approx  2bk\frac{e^{-k/(1-r)}}{(1-r)^2},\\
\phi'' &= b\left[\left(\frac{kr}{1-r^2}\right)''-2\left[\left(\frac{kr}{1-r^2}\right)'\right]^2 \text{tanh}\left(\frac{kr}{1-r^2}\right)\right]\text{sech}^2\left(\frac{kr}{1-r^2}\right) \\ &\approx b\left[\frac{4k}{(1-r)^3} - \frac{2k^2}{(1-r)^4}\right]e^{-k/(1-r)}.
\end{split}
\end{equation}
and
\begin{equation}
\begin{split}
S^2f &\approx - f'(1)S(1)^2(1-r)+...\\
(S^2f)' &\approx f'(1)S(1)^2 +...
\end{split}
\end{equation}
Using \eqref{eq:scalar} then yields
\begin{equation}
\frac{dV}{d\phi} = f'(1)\log^{2}\left(\frac{b-\phi}{2b}\right)\left(\log\left(\frac{b-\phi}{2b}\right)-1\right)\frac{b-\phi}{k}+\mathcal{O}[(b-\phi)^2],
\end{equation}
or
\begin{equation}
\begin{aligned}
V(\phi) = & \Lambda_{AdS} + \frac{f'(1)b}{4k}\left(\frac{b-\phi}{2b}\right)^2\left[-5+10\log\left(\frac{b-\phi}{2b}\right)\right. \\
& \left.-10\log^{2}\left(\frac{b-\phi}{2b}\right) + 4\log^{3}\left(\frac{b-\phi}{2b}\right)\right] +\mathcal{O}[(b-\phi)^3].
\label{eq:potential}
\end{aligned}
\end{equation}
\begin{figure}[t]
\centering
\includegraphics[scale=.4]{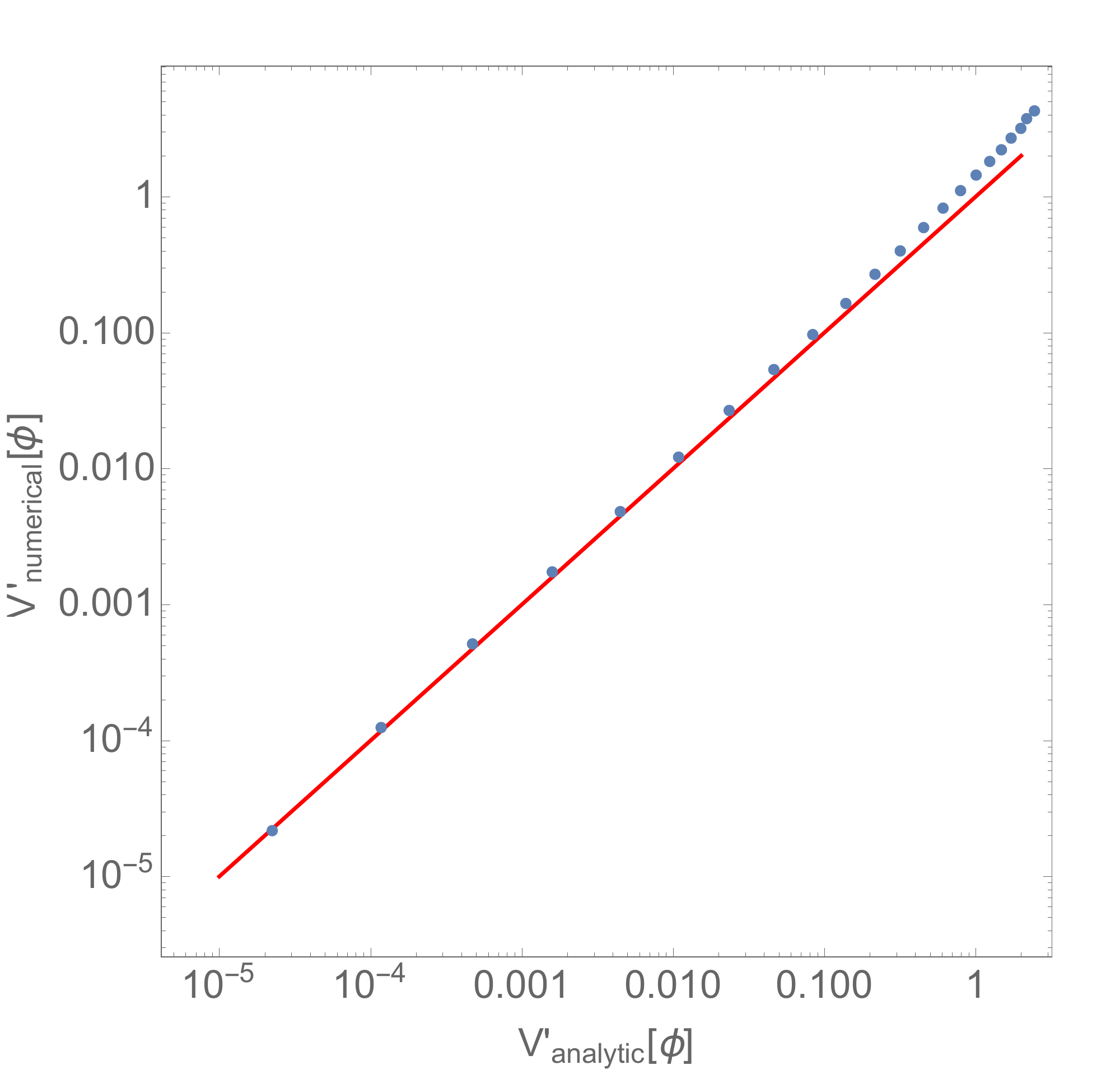}
\caption{A comparison of the numerical results (y-axis) for $\frac{dV}{d\phi}$ to analytic results (x-axis). We have plotted our result (blue) on a log-log plot against a line (red) with slope 1 to show agreement over 4 orders of magnitude. The values of $b$,$k$, and $r_h$ are the same as in fig. \ref{fig:solutions}.}
\label{fig:nvsa}
\end{figure}
So our potential is not smooth at its minimum.  Instead, $\frac{d^2V}{d\phi^2}$ has a logarithmic singularity, indicating that interactions remain important near the horizon.   As shown in figure \ref{fig:nvsa}, this result is consistent with our numerics.

We remark that the singularity in our potential is not just an artifact of our particular construction. Indeed, appendix \ref{sec:NoGo} demonstrates -- even when the requirement of a pure AdS-Schwarzschild exterior is dropped -- that time-independent spherically-symmetric wormholes cannot be sourced by scalar fields with smooth potentials.

\subsection{HRT entropies}
\label{sec:ThickHRT}

 Finally, we investigate the holographic HRT mutual information between the two boundaries of our smooth time-independent wormhole.  Here we consider the particular numerical solution displayed in figure \ref{fig:solutions}.  As in section \ref{sec:MIthin}, we begin by choosing $A_L,A_R$ to each be the northern hemisphere of the respective boundary at $t=0$.  Repeating the steps describes there, and since the solution is just AdS$_4$-Kruskal outside the horizon, we focus on the area $A_{\text{connected,\  inside}}$ of the surface defined by taking the equator of each sphere inside the horizon.  Interestingly, as in the thin-wall case, we find  $A_{\text{connected,\  inside}} > A_{EH}$ for all values of $k,b$ that we have explored -- and indeed even for other functional forms of $\phi(r)$ such as $\phi(r) = b\tanh\left(\frac{kr}{(r_h^2-r^2)^{c_2}}\right)^{c_1}$ with $c_1, c_2$ integer constants. So as in section \ref{sec:MIthin}  we conjecture that for general $A_L, A_R$ the HRT entropy $S_{A_LA_R}$ agrees with the $t \rightarrow +\infty$ limit of AdS-Kruskal and that, despite a likely instability analagous to that found for the domain wall solutions, in this sense our time-independent wormholes are already thermalized at any finite $t$. 
 
\begin{figure}[t]
\centering
\includegraphics[scale=.6]{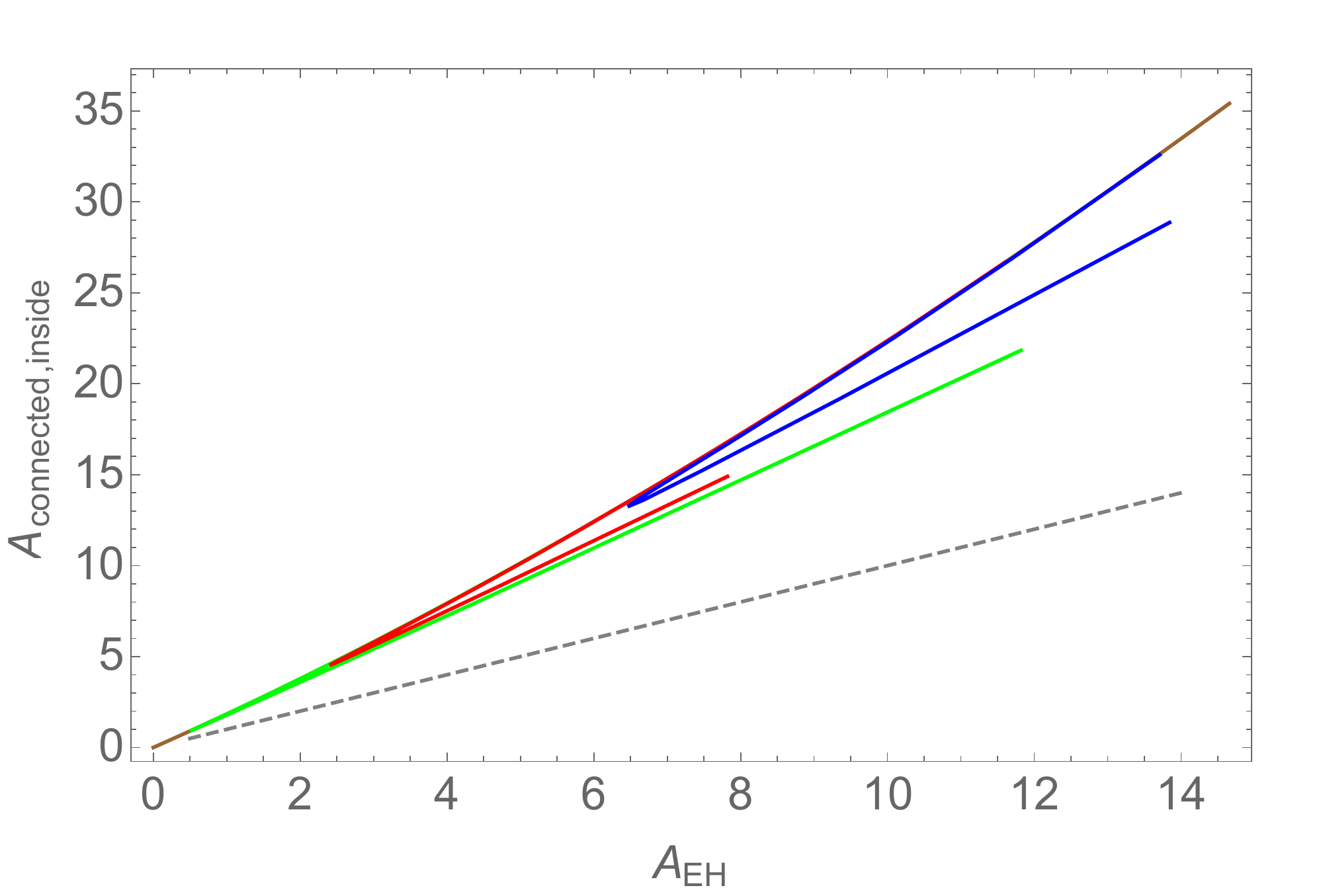}
\caption{A comparison of the areas $A_{\text{connected, \ inside}}$ of minimal surfaces inside the wormhole to the area $A_{EH}$ of the corresponding black hole. Using the profile in (\ref{eq:scalarprofile}), each curve corresponds to a fixed value of $b$ ($b=.87$ (green), $b=1$ (red), $b=1.2$ (blue)) while $k$ is varied from .2 to 2.1. For each $b$ there are two branches of solutions which join around $k\sim .7$. We have also plotted the $D=4$ solution from fig. \ref{fig: MutualInfo} in brown which is seen to coincide with the top branch of our solutions for each $b$; in particular, while the brown curve is hidden by the top branches of the colored curves across much of the figure, it remains visible at both the lower left and upper right ends. All solutions lie above the dashed line which plots $A_{\text{connected, \ inside}} =A_{EH}$, so the minimal surfaces are disconnected for hemispheres on the boundary of AdS.}
\label{fig:areas}
\end{figure}

 Typical results for $A_{\text{connected,\  inside}}$, $A_{EH}$ are shown in figure \ref{fig:areas} for the profile \eqref{eq:scalarprofile}.  One might expect that for large $k$ our smooth solutions approximate the thin-shell solutions of section \ref{sec:Thin}.    At least so far as these areas are concerned, the plot indicates that the agreement is already quite good for any $b$ at $k \sim 1$.  Indeed, different scalar profiles in this regime that lead to the same $A_{EH}$ also have nearly identical $A_{\text{connected,\  inside}}$.

\section{Discussion}
\label{sec:Disc}

We have constructed time-independent spherically symmetric AdS-wormholes sourced by both thin-shell domain walls and smooth scalar fields with potentials $V(\phi)$ that are $C^1$ but not $C^2$.  The time-translation in such spacetimes translates both wormhole mouths forward in time, instead of shifting them in opposite directions as in familiar AdS-Kruskal black holes. 
Interestingly, the results of figure \ref{fig:areas} indicates that, at least for some purposes, the thin-shell solutions become good approximations to the smooth solutions when the parameter $k$ in \eqref{eq:scalarprofile} satisfies $k \gtrsim 1.$

As shown in appendix \ref{sec:NoGo},
the non-smooth potential $V(\phi)$ is critical to the construction, as there there can be no precisely time-independent such solutions when the scalar potential $V$ is a smooth function of the scalar field $\phi$.  This feature may be related to the expectation that -- even when they exist -- the interior of such wormholes will be unstable.  The instability was identified explicitly in the thin-shell case.

Nevertheless, as discussed in sections \ref{sec:MIthin} and \ref{sec:ThickHRT}, at least for a large set of boundary regions the HRT entropies of boundary regions are already thermalized at any finite $t$ without the above instability having been triggered.  By this we mean that the result agrees with that obtained from familiar AdS-Kruskal in the limit $t\rightarrow \infty$.   This was shown in particular for many cases where the boundary region contains pieces on both boundaries so that the same result holds for cross-boundary mutual informations similar to those studied in \cite{Hartman:2013qma}.  Indeed, we conjecture that it holds for all such entropies and mutual informations.  Should one be able to find a stable version of our time-independent wormholes, a feature of this sort would be an interesting consistency check on whether dual gauge theory states thermalize in a universal way.

Such computations raise the question of whether our wormholes can have gauge theory duals in some version of gauge/gravity duality.    One question involves the dual description of the logarithm at the minimum of the potentials used in section \ref{sec:Thick}.  But leaving this aside for now, we might ask if our wormholes define stationary points of Euclidean path integrals in analogy with \cite{Maldacena:2001kr}.  At least in the thin-wall context, it is clear that the answer is negative.  Constructing a Euclidean thin-wall stationary point amounts to solving an ODE for the Euclidean motion of the wall within Euclidean AdS-Schwarzschild.  Since at $t=0$ the wall sits at \eqref{eq:sew} with zero velocity, it must do so for all Euclidean time.  But since shifting Euclidean time by half a period takes one to the opposite side of the Lorentzian horizon, this is incompatible with the requirement that the wall exist only inside the wormhole and not outside.    It would be interesting to determine whether a similar argument applies to our smooth wormhole solutions built from non-smooth potentials.

Finally, we briefly mention the recent discussions of the possible role of complexity in gauge/gravity duality in \cite{Stanford:2014jda,Roberts:2014isa,Brown:2015bva,Brown:2015lvg} and the conjectures that gauge-theory complexity is related either to the volume of maximal slices or the action of certain regions in the bulk geometry.  In our case, even the renormalized volume of a maximal slice that extends from one boundary to the other is strictly infinite.  Although the renormalized volume of the $t=0$ slice will be finite, for maximal slices there is no analogue of the argument in footnote \ref{VH}.  Indeed, in our case it is clear that a surface of arbitarily large renormalized volume can be obtained by following an orbit of the Killing field in the regions of figure \ref{fig:Conformal} (right) in which the time-translation Killing field is spacelike.  In particular, the volume of such surfaces grows without bound as the surface nears the topmost point of the dotted line in \ref{fig:Conformal} (right).  Similarly, the action of the spacetime region inside the wormhole (say, defined as in \cite{Lehner:2016vdi}) should diverge due to the required integral over time and the (non-compact) time-translation symmetry.  Interestingly, assuming the wormhole to be unstable as in section \ref{sec:Thin} and choosing a perturbation that collapses the interior even at a very late time would result in finite actions and volumes of maximal surfaces at any given time $t$, though the resulting breaking of time-translation symmetry would also cause these quantities to grow with time.  Indeed, at late times the growth in such quantities should be dominated by the region near the outermost horizon and so will proceed precisely as in AdS-Kruskal.  In contrast, with an instability that causes the wormhole interior to expand both the relevant actions and volumes of maximal slices will continue to diverge.  It would be interesting to understand better the meaning of such divergences in the context of the conjectures of \cite{Stanford:2014jda,Roberts:2014isa,Brown:2015bva,Brown:2015lvg}.

\section*{Acknowledgements}
We thank Gary Horowitz, Henry Maxfield, Brian Swingle, and Ying Zhao for useful discussions.  DM and ZF were supported in part by the Simons Foundation and by funds from the University of California. EM was supported in part by NSF grant PHY-1504541.

\appendix

\section{No solutions with smooth scalar potentials}
\label{sec:NoGo}

This appendix shows that spherically symmetric time-independent wormholes with a $Z_2$ reflection cannot be sourced by scalar fields with smooth potentials, and thus that the singular potential found in section \ref{sec:Thick} is not an artifact of our particular construction.  Our argument closely follows the work of Bekenstein \cite{Bekenstein:1995un} constraining black holes with scalar hair, though we have rephrased much of the proof in terms of manifestly covariant quantities like the expansion of radial geodesics.  We allow a general scalar action of the form
\begin{equation}
\mathcal{L} = \left[\frac{g^{ab}}{2}M_{AB}(\{\phi^A\})\;\partial_a\phi^{A}\;\partial_b\phi^{B}+ V(\{\phi^A\}) \right] \sqrt{-g}
\end{equation}
with positive definite $M_{AB}$.  Such fields in particular respects the null energy condition (NEC), which states that the stress tensor $T_{ab}$ satisfies $T_{ab}k^a k^b \ge 0$ for all null $k^a$.

We again describe the spacetime using the metric \eqref{eq:wmet}, taking $r=0$ at the surface invariant under the $Z_2$ symmetry.  We also assume the scalar fields to share the symmetries of the spacetime so that they depend only on the coordinate $r$. As a result, covariance and the definition $T_{ab} = -\frac{2}{\sqrt{-g}}\frac{\delta S}{\delta g^{ab}}$ require
\begin{eqnarray}
\label{eq:stress}
- T_t^{\;\;t} = - T_{\Theta}^{\;\;\Theta} = \mathcal{E} = \frac{f}{2}M_{AB}(\{\phi^A\})\;\partial_r\phi^{A}\;\partial_r\phi^{B} + V(\phi) , \\
T^r_r = \frac{f}{2}M_{AB}(\{\phi^A\})\;\partial_r\phi^{A}\;\partial_r\phi^{B} - V(\phi)
\end{eqnarray}
where $\mathcal{E}$ is the Lagrangian density for the scalars and $T_{\Theta}^{\;\;\Theta}$ is the same for all angular coordinates.  Furthermore, the scalar equation of motion takes the form
\begin{equation}
\label{eq:multiphi}
\partial_r \left(M_{AB} f S^{d-1} \partial_r \phi^B\right) = \frac{\partial V}{\partial \phi^A}.
\end{equation}
The Einstein equations remain as in \eqref{eq:Einstein} with the substitution of \eqref{eq:stress}.

Our argument begins not in the wormhole itself, but in the region outside the horizon.  Here we recall that the null convergence condition (a consequence of the Einstein equations and the null energy condition) requires the size of the $(d-1)$ spheres to increase monotonically from the horizon to the conformal boundary.  The argument is simply that spherical symmetry prevents any outgoing sphere of light rays from forming a caustic before reaching the conformal boundary, and that a standard calculation shows that such null congruences have vanishing expansion $(\theta =0)$ at the AdS boundary.  The Raychaudhuri equation thus forbids them from having $\theta < 0$ at any $r$ outside the horizon and thus implies monotonicity of $S(r)$.  In the same way, since $S'(r)=0$ at $r=0$, the sphere size $S$ must increase monotonically as one moves from a horizon toward $r=0$.  So the horizon $r=r_h$ must be a local minimum of $S(r)$.

In contrast to section \ref{sec:Thick}, we now wish to take $V(\phi)$ to be a fixed smooth function of $\phi$ and to solve \eqref{eq:multiphi}, \eqref{eq:Einstein} to generate the spacetime.  We seek solutions with $S\neq 0$, so the only singular points of this system of ODEs occur when $f=0$; i.e., at the horizon.  In order for this to be a smooth Killing horizon with finite surface gravity, both $f$ and must be a smooth function of $r$ at $r_h$.  Thus as usual $r-r_h$ is quadratic in the proper distance $s-s_h$ from the horizon, and in fact  $(r-r_h)$ may be expanded in even powers of $(s-s_h)^2$.  Smoothness of the geometry then requires that $f$ also have an asymptotic series expansion about the horizon, and that this expansion is even.

This suggests that the entire solution will be symmetric about the horizon.  Given that on one side we allow no further horizons between $r_h$ and the AdS boundary, such a symmetry would forbid the desired wormhole from being present on the other.  Indeed, we will show below that smoothness of $V(\phi)$ prohibits any breaking of this symmetry by $S, \phi^A$, or by effects vanishing faster than any power of $r-r_h$ and thus forbids smooth time-independent wormholes.

We begin with perturbative effects.  Having shown above that $r_h$ is a minimum of $S$, smoothness requires $S = S(r_h) + \frac{dS}{dr} |_{r_h}(s-s_h)^2 + O((s-s_h)^3)$ with $\frac{dS}{dr}$ finite at $r_h$. The scalar equation of motion \eqref{eq:multiphi} then also forces $\frac{d\phi^B}{dr}$ to remain finite and in fact constrains its value.  Repeated differentiation of \eqref{eq:multiphi} and the first equation of \eqref{eq:Einstein} then guarantee that all $r$-derivatives of $S, \phi^B$ are finite at $r_h$ as well so that they also admit well-defined asymptotic series expansions involving only even powers of $s-s_h$.

We now consider possible non-perturbative effects.  In particular, suppose that two solutions $(f_1,S_1, \phi^A_1)$ and $(f_2,S_2, \phi^A_2)$ have identical asymptotic expansions about the horizon.  Near $r_h$, we may then expand our ODEs in powers of $\Delta f = f_1 - f_2, \Delta S = S_1 -S_2$, and $\Delta \phi^A = \phi^A_1 -  \phi^A_2$.  And close enough to $r_h$, to good approximation we may truncate this expansion to first order and neglect $f \Delta \partial_r \phi^A$ relative to  $\partial_r \phi^A$.  Doing so results in a linear system of ODEs for
$\Delta f, \Delta S$, and $\Delta \phi^A$ with smooth non-vanishing coefficents; in particular, the ODE resulting from \eqref{eq:multiphi} is only of first order.  The boundary condition that $\Delta f, \Delta S$, and $\Delta \phi^A$ all vanish at $r_h$ thus requires them to vanish everywhere.  We have thus shown that solutions of our ODEs are uniquely determined by their power series expansion near $r_h$ for smooth $V(\phi)$.  As a result, smooth $V(\phi)$ requires a $Z_2$ symmetry about any smooth bifurcate Killing horizons an forbids the desired time-independent wormholes.

\bibliographystyle{JHEP}
	\cleardoublepage
\phantomsection
\renewcommand*{\bibname}{References}

\bibliography{wormhole}

\end{document}